\documentclass[twocolumn,showpacs,superscriptaddress,preprintnumbers,amsmath,amssymb,epsfig,floatfix,prb]{revtex4}
\usepackage{amssymb}
\usepackage{epsfig}
\usepackage{calrsfs}
\DeclareMathAlphabet{\pazocal}{OMS}{zplm}{m}{n}
\usepackage{graphicx}
\usepackage{dcolumn}
\usepackage{color}
\usepackage{bm}

\begin{document}
\title {Magnetism in stoichiometric and off-stoichiometric MnO clusters: \\ Insights from {\it ab initio} theory}	
\author{Shreemoyee Ganguly}
\altaffiliation{Corresponding author: ganguly.shreemoyee@gmail.com}
\affiliation{Division of Materials Theory, Department of Physics and Astronomy, Uppsala University, Box 516, SE-75120 Uppsala, Sweden.}
\author{Mukul Kabir}
\affiliation{Indian Institute of Science Education and Research, Pune 411008, India }
\author{Carmine Autieri}
\affiliation{Division of Materials Theory, Department of Physics and Astronomy, Uppsala University, Box 516, SE-75120 Uppsala, Sweden.}
\author{Biplab Sanyal}
\affiliation{Division of Materials Theory, Department of Physics and Astronomy, Uppsala University, Box 516, SE-75120 Uppsala, Sweden.}

\date{\today}
\begin{abstract}
{We study the composition dependent evolution of geometric and magnetic structures of MnO clusters within density functional theory. A systematic and extensive search through the potential energy surface is performed to identify the correct ground state, and significant isomers. We find that the magnetic structures in these MnO clusters are complex, which has been explained using the intrinsic electronic structure of the cluster, and analyzed using} model Hamiltonian with parameters obtained from maximally localized Wannier functions. The calculated vertical displacement energies of off-stoichiometric MnO clusters compare well with the recent experimental results.  {Interestingly, the charged state of the cluster strongly influences the geometry and the magnetic structure of the cluster, which are very different from the corresponding neutral counterpart. Further, the importance of electron correlation in describing simple Mn-dimer and MnO clusters has been discussed within Hubbard model and hybrid exchange-correlation functional.}  
\end{abstract}

\maketitle

\section{Introduction}
The primary interest in atomic clusters today stems from the fact that their properties are significantly different from that of the corresponding bulk.~\cite{Lucas,Lopez,Li,Billas} Transition metal oxide, mainly MnO, is extremely important as catalysts in many industrial chemical applications.~\cite{Kung} The small stoichiometric (MnO)$_x$ clusters ($x$=1--8) are structurally and magnetically very different from the bulk counterpart.~\cite{MnO} Surprisingly, the clusters containing up to five MnO units adopt two dimensional structures. Moreover, similar to the bulk MnO, the Mn moments in these clusters are antiferromagnetically coupled, which is debated in literature.~\cite{Jena1} However, the recent negative-ion photoelectron spectroscopy measurements have provided some interesting and intriguing results.~\cite{MnO-recent} For example, It has been shown that the Mn$_3$O$^-$ cluster is three dimensional, although the charge neutral stoichiometric cluster (MnO)$_3$ was predicted to be planar.~\cite{MnO} In addition, the anionic cluster Mn$_2$O$^-$ is predicted to be ferromagnetic,~\cite{Jena4} and Mn$_4$O$^-$ is found to have a substantial finite moment.~\cite{MnO-recent}  In contrast, the corresponding stoichiometric neutral clusters (MnO)$_2$ and (MnO)$_4$ are antiferromagnetic with zero net moment.~\cite{MnO}

Moreover, in contrast to the small pure Mn$_x$ clusters ($x$=2--4), which are found to be ferromagnetic,~\cite{Kabir2006} the stoichiometric (MnO)$_x$  ($x$=1--8) clusters are predicted to be antiferromagnetic.~\cite{MnO}  Thus, a magnetic transition is expected due to oxidation.  In addition, while the pure Mn$_4$ and Mn$_5$ are three dimensional,~\cite{Kabir2006,MnO} the (MnO)$_4$ and (MnO)$_5$ clusters are found to be two dimensional.~\cite{MnO} Thus, in addition to a magnetic transition, there exists a structural transition upon oxidation. Therefore, tracing the path of these magnetic and structural transitions is intriguing. This would also provide with a tool to control the magnetism in these clusters by tuning oxygen concentration. {Although, the small anionic off-stoichiometric clusters have been studied recently,~\cite{MnO-recent}  the microscopic origin of these complex geometric and magnetic evolution is not yet understood.} 
 
{With these motivations, here we study the evolution of geometric and magnetic structures of MnO clusters with varied composition and charged state, within density functional theory (DFT). The ground state, and the corresponding isomers are predicted via an extensive search through the potential energy surface. Moreover, the microscopic origin of such complex evolution is explained via the intrinsic electronic structure of the clusters. The results are compared with the experimental results.~\cite{MnO-recent} Further, we explain the necessity of strong correlation to describe the Mn-dimer via DFT corrected with on-site Coulomb interaction, hybrid functional, and model Hamiltonian approach. On similar lines, the effect of correlation and thus the necessity of using hybrid exchange-correlation functional to describe the MnO clusters have been debated in literature.~\cite{kino} In this regard, we calculate the vertical displacement energies (VDE) for off-stoichiometric clusters using both conventional Perdew-Burke-Ernzerhof (PBE) and hybrid exchange-correlation functionals, which are compared with the available experimental results.~\cite{MnO-recent} Moreover, we find that both the geometric and magnetic structures are strongly influenced by the charged state of the cluster. Such prediction was not possible in the previous calculations, where the ground state search for the neutral clusters was biased by the results for the corresponding anionic clusters.~\cite{MnO-recent}}   

The paper evolves as the following. In Section II, we discuss the computational details. In Section III-A, we revisit the stoichiometric clusters, with an emphasis on the closely lying isomers. The widely debated Mn-dimer is discussed in the next section using various theoretical hierarchy, which is inadequately described within conventional DFT.~\cite{Kabir2006} The results are interpreted using model Hamiltonian.  In Section III-C, the neutral and anionic non-stoichiometric clusters are reported along with the corresponding magnetic and structural transition. In the next section, we discuss the calculated VDE using different exchange-correlational functionals, and make a detailed comparison with the experimental results. Finally in the Section IV, we summarise our results, and conclude. 
 
\section{Computational Details}               
The calculations are performed using density functional theory based pseudopotential plane wave method.~\cite{Kresse1999} Here used the projector augmented wave method,~\cite{Blochl1994} and the Perdew-Burke-Ernzerhof exchange-correlation functional~\cite{Perdew1996} for the spin-polarized generalized gradient correction as implemented in the VASP code.\cite{vasp} We have repeated few calculations using the PBE0 hybrid functional, where the exchange-correlation energy functional is written as, 
\begin{eqnarray}
{E_{xc}^{\rm PBE0}} = {\frac{1}{4}E_x}+{\frac{3}{4}E_x^{\rm PBE}}+{\frac{1}{4}E_c^{\rm PBE}}, 
\end{eqnarray}
where $E_{x}$ is the exact exchange, and $E_x^{\rm PBE}$ and $E_c^{\rm PBE}$ are the GGA-PBE exchange and correlation energies, respectively. Calculations using the on-site Coulomb interaction has been done within the DFT+$U$ approach as described by Dudarev {\it et al}.~\cite{Dudarev} The 3$d$ and 4$s$ electrons of Mn, and 2$s$ and 2$p$ electrons of O are treated as valence electrons. The wave functions are expanded in a plane wave basis set with 270 eV kinetic energy cut-off. Reciprocal space integrations are carried out at the ${\Gamma}$ point. 

For all the clusters studied here, stoichiometric (MnO)$_x$ ($x$=1--8) and off-stoichiometric Mn$_x$O$_y$ ($x$=2--4 and $y<x$) clusters, we considered many different initial geometric structures through spin-polarized Born-Oppenheimer molecular dynamics (BOMD) simulation.  The complex potential energy surface of the cluster is extensively sampled within canonical ensemble using Nos\'e-Hoover thermostat.\cite{Nose}  Starting from a high symmetry structure [for example a cubic structure similar to the core of Mn$_{12}$-molecular magnet~\cite{Nature.365} for the (MnO)$_4$ cluster] we have heated the clusters to 2000K (above the melting temperature of bulk MnO), and performed BOMD simulations for 6--10 ps.  At elevated temperature, the cluster evolves through various geometry.
Carefully studying the structural evolution at this elevated temperature, we picked many structures with different symmetry. All these structures were further optimized considering all possible spin multiplicities, until all the forces are less than a threshold value of 5 meV/\AA. This ensures the robustness of the ground state search, which was previously employed by us.~\cite{MnO} It should be mentioned here that the calculations are done within the collinear spin assumption.

\begin{table}[!t]
\caption{\label{tab1} Binding energies per MnO-unit ($E_B$/MnO), total magnetic moment of the cluster $\mu_{\rm tot}$, and the total Mn$\rightarrow$O charge transfer (CT) for stoichiometric (MnO)$_x$ clusters. The energy difference between the AFM ground state and the most stable FM isomer $\Delta E$, and the total hybridization index $\mathcal{H}$ for 3D and 2D structures are also shown.}
\begin{tabular}{lcccccc}
\hline
\hline
$x$  &  $E_B$ & $\mu_{\rm tot}$ & Total Mn$\rightarrow$O  & $\Delta E$ & $\mathcal{H}^{3D}$&$\mathcal{H}^{2D}$ \\
   &  (eV/MnO) & ($\mu_B$) & CT ($e$) & (eV/MnO) & & \\  
\hline
\hline
2 & 7.34 & 0 & 1.2 & 0.17 & $-$ & 2.26\\
3 & 8.25 & 5 & 1.3 & 0.14 & $-$ & 3.84\\
4 & 8.58 & 0 & 1.3 & 0.18 & 4.91 & 5.68\\
5 & 8.64 & 5 & 1.2 & 0.24 & 6.58 & 7.01\\
6 & 8.82 & 0 & 1.3 & 0.13 & 8.04 & 7.87\\
7 & 8.89 & 5 & 1.3 & 0.26 & 9.01 & 8.84\\
8 & 8.99 & 0 & 1.3 & 0.21 & 10.64 & $-$\\
\hline
\end{tabular}
\end{table}

For stoichiometric clusters,  the binding energy per MnO-unit ($E_{b}$) is defined as,
\begin{equation}
E_b[{\rm (MnO)}_{x}]=\frac{1}{x}[xE({\rm Mn})+xE({\rm O})-E({\rm (MnO)}_{x})],
\end{equation}
where $x$ is the number of MnO-units in our cluster, $E[{\rm (MnO)}_x$], $E({\rm Mn}$), and  $E({\rm O})$ are the total energies of (MnO)$_{x}$ cluster, and an isolated Mn and O atom, respectively. For a given $x$, the structure with the highest binding energy is considered to be the `ground state'. 

For off-stoichiometric Mn$_x$O$_y$ clusters, the binding energy per atom is defined as,
\begin{equation}
E_b({\rm Mn}_{x}{\rm O}_{y})=\frac{1}{n}[xE({\rm Mn})+yE({\rm O})-E({\rm Mn}_{x}{\rm O}_{y})],
\end{equation}
where $n=x+y$ is the total number of atoms in the cluster, and $E({\rm Mn}_{x}{\rm O}_{y}$) is the total energy of ${\rm Mn}_{x}{\rm O}_{y}$ cluster. The local magnetic moment $\mu_{\rm X}$ at X-atom is calculated as,
\begin{equation}
\mu_{\rm X}=\int_0^R[\rho_{\uparrow}({\mathbf r})-\rho_{\downarrow}({\mathbf r})]\,d\mathbf{r}
\end{equation}
where $\rho_{\uparrow}({\mathbf r})$ and $\rho_{\downarrow}({\mathbf r})$ are spin-up and spin-down charge-densities, respectively, and $R$ is the radius of the sphere centred on the atom X, which depends on the atom type, Mn/O.~\cite{Note1} 

\begin{figure*}[t]
\includegraphics[scale=0.5]{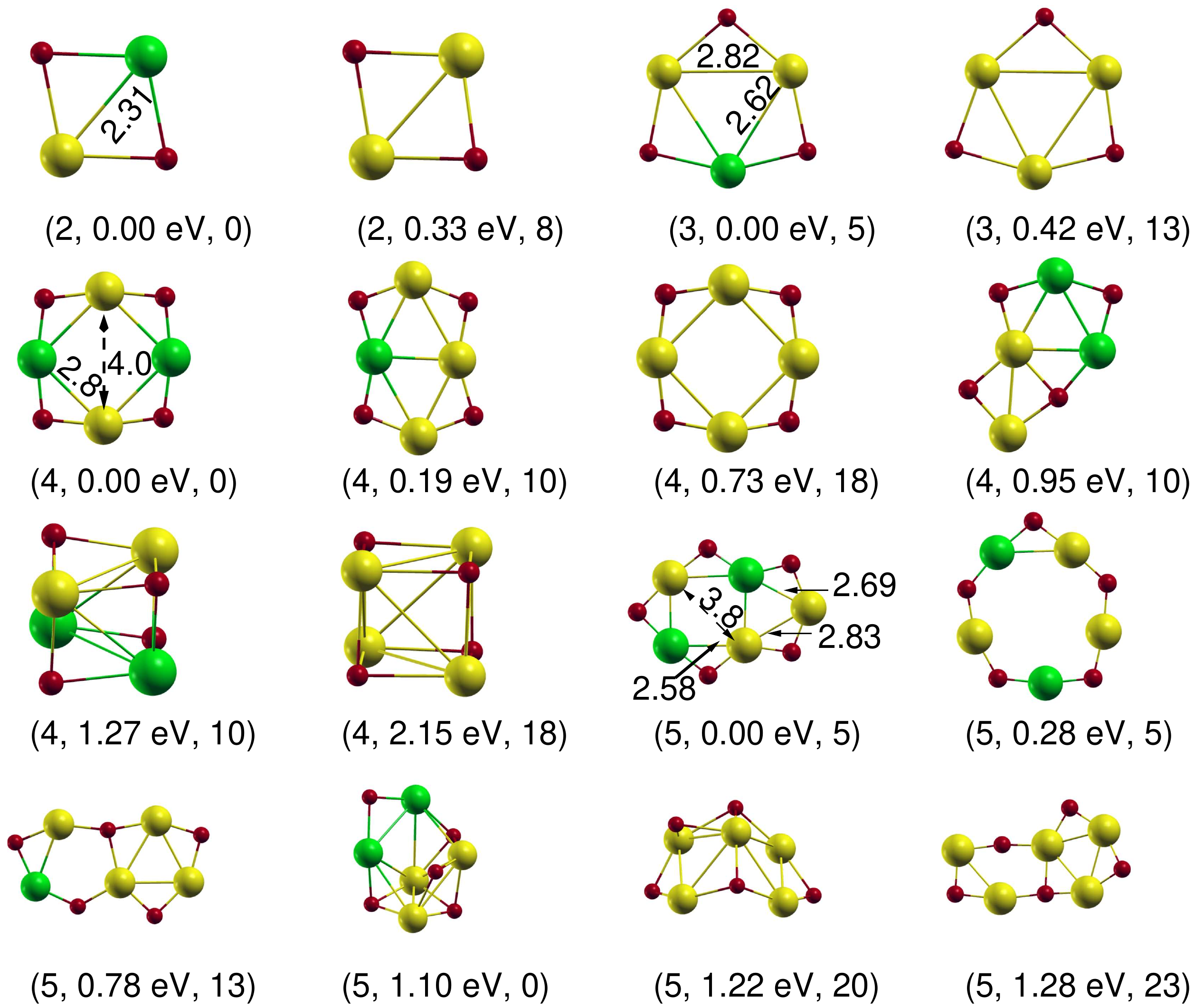}
\caption{\label{fig:trimer}(Color online) Ground state and the significant isomers of stoichiometric (MnO)$_x$ ($x$=2--5) clusters. Smaller (Red) spheres represent O atoms, and larger yellow (green) spheres represent Mn$_{\uparrow}$ (Mn$_{\downarrow}$) atoms. The numbers in the parenthesis represent the cluster size [$x$ in (MnO)$_x$],  energy relative to the corresponding ground state $\Delta E$, and the total magnetic moment, respectively. Ground states are obtained via an extensive search through the potential energy surface. The Mn-Mn distances are shown for the ground states. {In these stoichiometric clusters, the AFM coupled Mn atoms are closer than those coupled ferromagnetically}.}
\end{figure*}

The orbital hybridization can be quantified, and would be useful to explain the cluster morphology. This was applied earlier to explain the 2D nature of the gold clusters.~\cite{Hakkinen}  We calculate the $k$-$l$ hybridization index, 
\begin{equation}
\mathcal{H}_{kl} = \sum_I \sum_i w_{ik}^I w_{il}^I, 
\label{equation}
\end{equation}
where $k$, $l$ are the orbital indices,  $w_{ik}^I (w_{il}^I)$ is the square projection of the $i$-th Kohn-Sham orbital on to the $k$ ($l$) spherical harmonics centered at atom $I$ and integrated over a sphere.  Note that the spin index is inherent. However, unlike gold clusters, in a system with active $p_{\rm O}$ electrons, in addition to the $s$-$d$ hybridization ($\mathcal{H}_{sd}$), the $\mathcal{H}_{pd}$ and $\mathcal{H}_{sp}$ would also play an important role in determining the dimensionality. 

We calculate the electron hopping parameters using the Slater-Koster interpolation scheme based on the localized Wannier functions.~\cite{Slater54} Such approach is applied to determine the real space Hamiltonian in the $d$-like and $p$-like Wannier function basis.  After obtaining the Bloch bands in density functional theory, the Wannier function are constructed using the WANNIER90 code.\cite{Mostofi08} Starting from an initial projection of $d$-like atomic wavefunctions centered on Mn sites, and $p$-like atomic wavefunctions centered on O sites, we obtain the Wannier basis.

\begin{figure*}[t]
\includegraphics[scale=0.5]{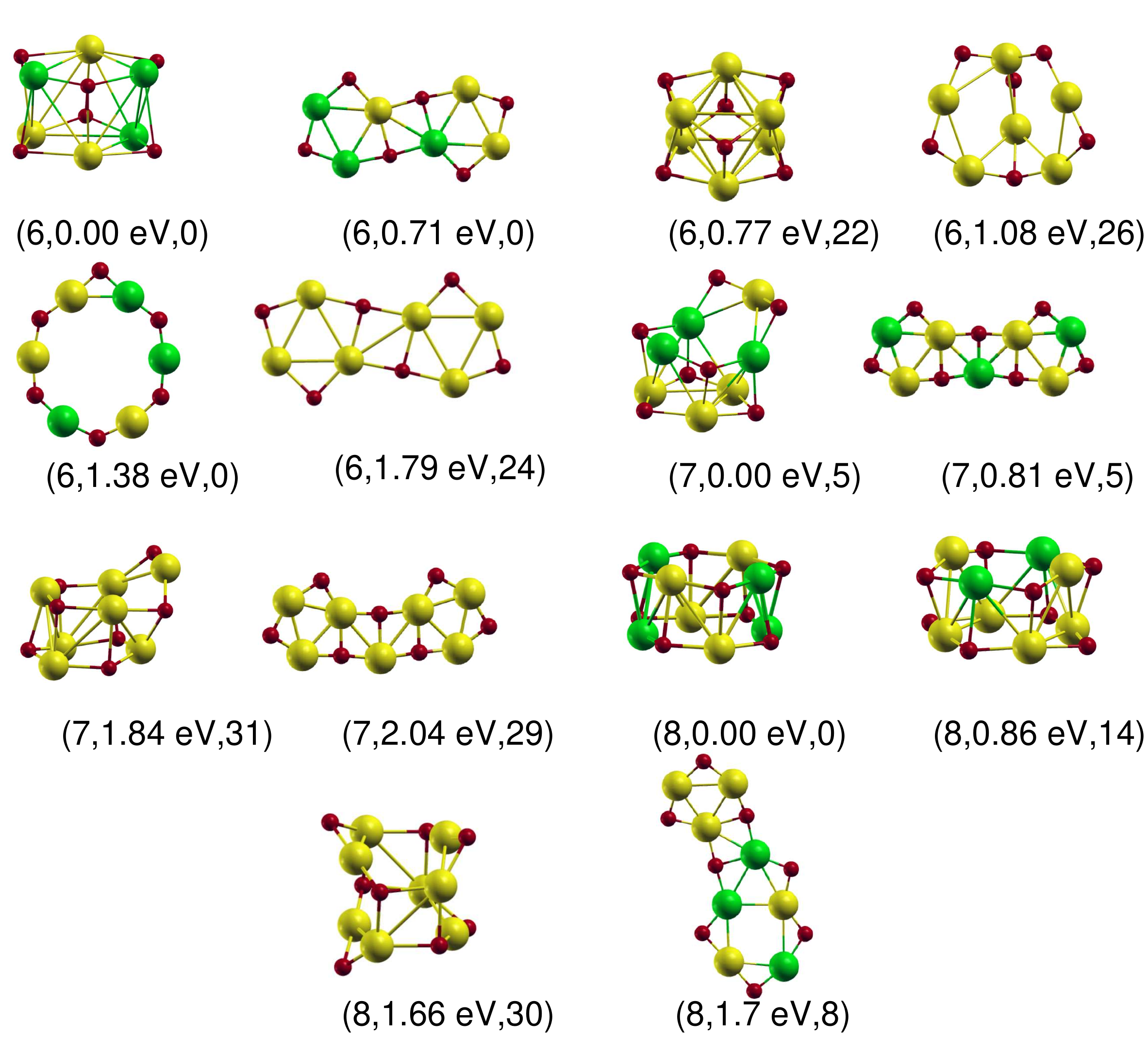}
\caption{\label{fig:five}(Color online) Ground state, and the significant isomers of stoichiometric (MnO)$_x$ ($x$= 6--8) clusters. We have used the same convention as in Fig~\ref{fig:trimer}.  A 2D$\rightarrow$3D structural transition takes place for (MnO)$_6$, which can be explained by the Mn-O hybridization. The energy difference between the 3D ground state and the most likely 2D structure is 0.71 eV for (MnO)$_6$, and further increases with increasing clusters size. The ground states are found to be AFM, while the most likely FM structure is much higher in energy.}
\label{fig:st68}
\end{figure*}

\section{Results and Discussions}
\subsection{\label{stoich} Stoichiometric MnO Clusters Revisited}
Although we have earlier discussed the ground states of stoichiometric MnO clusters,~\cite{MnO} here we begin our discussion including the corresponding significant isomers. {The information of isomers is very important as the cluster experiments are done at finite temperature, and thus the experimental cluster beam may have a mixture of energetically close isomers.} The binding energy, magnetic moment, energy difference between the ground state with the closest ferromagnetic isomer, and hybridization indices are summarized in Table \ref{tab1}. The cluster geometries are shown in Fig.~\ref{fig:trimer}. 

The intrinsic Mn-Mn coupling in (MnO)$_{2}$ is found to be antiferromagnetic in the ground state, while the closest ferromagnetic (FM) isomer is energetically 0.33 eV higher. For (MnO)$_3$ cluster, the ground state is a ring of alternating Mn and O atoms, which is in agreement with the previous prediction.~\cite{Jena1}  In this structure, Mn atoms which are  antiferromagnetically coupled are closer (2.60 \AA) compared to the FM Mn$_3$ trimer (2.74 \AA).~\cite{Kabir2006}  For such oxide clusters the exchange pathways are nontrivial with the competing Mn-Mn direct exchange, and Mn-O-Mn superexchange. We will discuss this in a later section.  The closest FM isomer is 0.42 eV above in energy. The (MnO)$_4$ has a two-dimensional structure, where Mn and O atoms alternate. Here again the adjacent Mn atoms are antiferromagnetically coupled. The closest isomer has a ferrimagnetic (FE) structure due to the reorientation of Mn-Mn bonds (Fig\ref{fig:trimer}). Interestingly, the predicted geometric and magnetic structures are very different from the previous theoretical predictions.~\cite{Jena1,Jena2,Pederson}  The closest 3D isomer is energetically very high (1.27 eV).  The ground state structure of (MnO)$_5$ can be viewed as a combination of (MnO)$_3$ and (MnO)$_4$ geometries, with AFM (FM) Mn-Mn coupling for shorter (longer) Mn-Mn bonds.  First two isomers are also found to be two-dimensional. A three-dimensional isomer is the third isomer, which lies 1.10 eV above the ground state. The Mn-core of this trigonal bipyramidal isomer resembles with the pure Mn$_5$ ground state. 

We observe an emergence of three-dimensional ground state for (MnO)$_6$ cluster, which can be viewed as two (MnO)$_3$ rings stacked together (Fig.~\ref{fig:st68}), and was suggested by Ziemann and Castlemann.\cite{Ziemann} In each of these rings, AFM coupled Mn atoms are closer. A two-dimensional FM isomer lies 1.08 eV above in energy, whereas the closest ferromagnetic isomer lies 0.77 eV above the ground state. These observations are strikingly different from previous theoretical predictions.\cite{Jena1,Jena2} Although the presumed cubic structures are very different from the present ground state structures, similar AFM ground state has been earlier reported.~\cite{Han} The ground state of (MnO)$_7$ is found to be a fused (MnO)$_3$/(MnO)$_4$ geometry in three dimension. The closest planar isomer is 0.81 eV higher in energy, and the closest FM isomer is 1.84 eV above. Planer (MnO)$_4$ serves as the building block for (MnO)$_8$ cluster, where two (MnO)$_4$ blocks are stacked in three-dimension for the ground state (Fig.~\ref{fig:st68}). Similar to other clusters, the AFM coupled Mn atoms are closer than that of the FM coupled ones in each of these (MnO)$_4$ units. Although, the AFM ground state is in agreement with the earlier prediction, the predicted geometry is very different as they considered a particular structure only.~\cite{Han} In contrast with the previous prediction,~\cite{Jena1,Jena2} we do not observe any magnetic bistability for this cluster. Rather, using an extensive potential energy surface scanning technique, we find the AFM ground state to lie much lower in energy (1.66 eV) than the FM state, while the closest ferrimagnetic solution is 0.86 eV higher in energy.  

\begin{figure}[!t]
\includegraphics[width=6cm,  keepaspectratio]{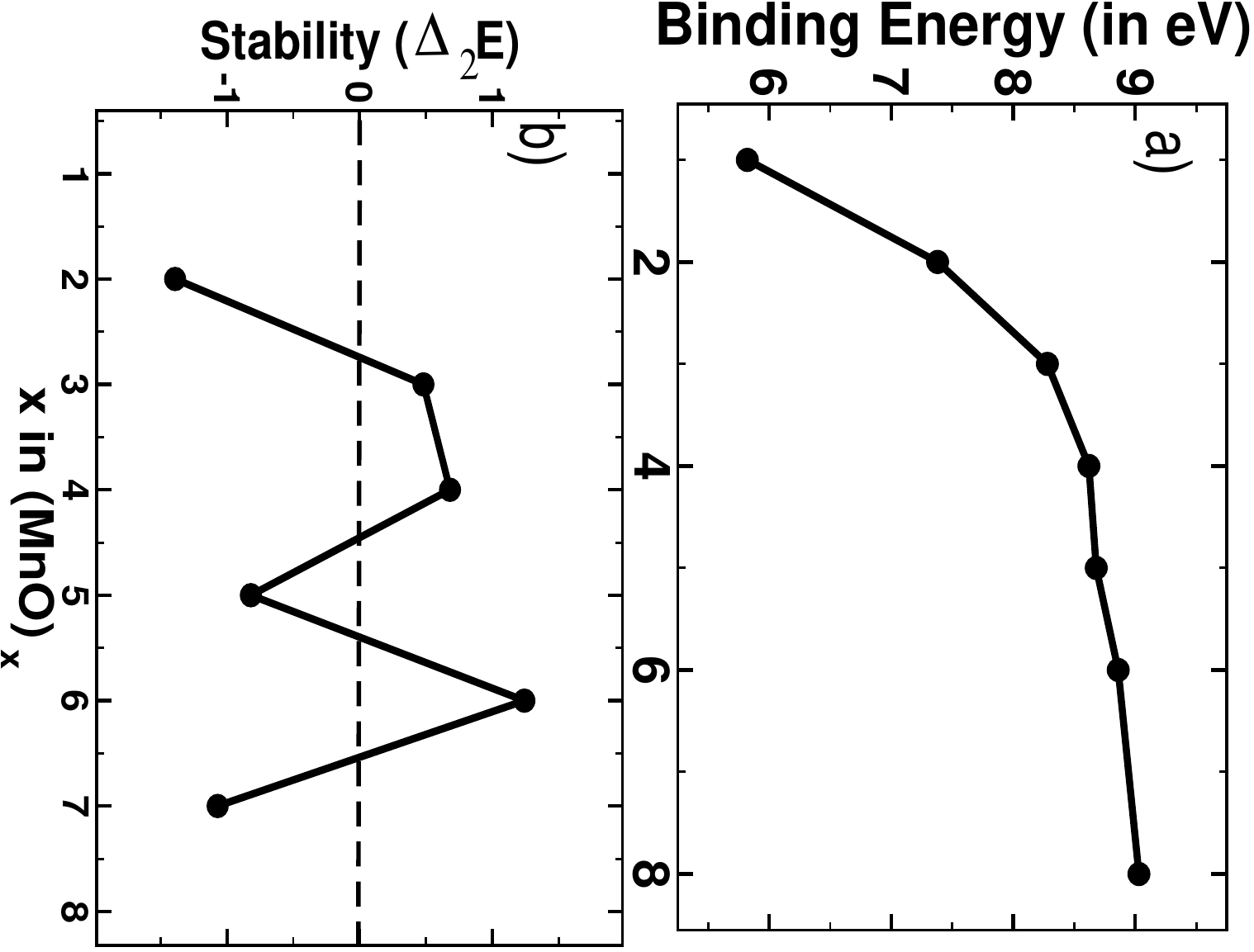}
\caption{\label{fig:be} (a) Binding energy increases with increasing MnO-unit. The rate of increase in binding energy slows down with the increase in $x$,  and attains $\sim$ 9 eV/MnO-unit, which is close to the experimental bulk value of 9.5 eV/MnO-unit.~\cite{Barin} (b) Local stability $\Delta_2E$ as a function of $x$ shows that the (MnO)$_3$, (MnO)$_4$ and (MnO)$_6$ clusters are particularly stable, which is in agreement with experimental results.~\cite{Ziemann}}
\end{figure}

Next, we discuss the trend in the binding energy for these stoichiometric clusters. The coordination number increases with the cluster size, and hence the binding energy [Fig.~\ref{fig:be}(a)], which saturates at $\sim$ 9 eV/MnO-unit. This is close to the experimental bulk value of 9.5 eV/MnO-unit.~\cite{Barin} To evaluate the local stability of these stoichiometric clusters, we calculate the second difference in energy,
\begin{equation*}
 \Delta_2E =E({\rm MnO})_{x-1} +E({\rm MnO})_{x+1}  - 2E({\rm MnO})_{x}. 
 \end{equation*}
Positive $\Delta_2E$ indicates the greater stability of (MnO)$_3$, (Mn4O)$_4$ and (MnO)$_6$ clusters [Fig.~\ref{fig:be}(b)], which is in agreement with the previous theoretical and experimental predictions.~\cite{Jena1,Ziemann}   Due to the higher stability of (MnO)$_3$ and (Mn4O)$_4$ clusters, they are found to be the building blocks for other clusters (Fig.~\ref{fig:trimer} and Fig.~\ref{fig:st68}). The (MnO)$_6$ cluster has greater stability as this is made of two  (MnO)$_3$ clusters staked in three-dimension.  These observations are in good agreement with the experimental prediction through mass spectroscopy, which concludes that the [(MnO)$_3$]$_n$ clusters, with $n$=1, 2, 3 and 4 are more abundant. In addition, we also predict greater stability for (MnO)$_4$, and thus we find it to serve as the building blocks for  (MnO)$_7$ and (MnO)$_8$ clusters (Fig.~\ref{fig:st68}).   

The MnO cluster has a net moment of 5$\mu_B$, which is in good agreement with the experiment.~\cite{Ryan} For all the clusters studied here, the majority of this moment is localized on the Mn atom as theoretically predicted earlier.~\cite{MnO,Jena3,Pederson} Similar to the bulk counterpart,~\cite{Towler} the Mn-Mn coupling is antiferromagnetic for all the stoichiometric clusters studied here.  While for clusters with even number of MnO units the net moment is 0$\mu_B$,  those with the odd number of MnO units have a total moment of 5$\mu_B$ due to unequal number of up and down atoms. As we have discussed earlier, for all the clusters ferromagnetic isomers are well above in energy. Such AFM Mn-Mn coupling is in agreement with the previous calculation for cubic (MnO)$_x$ clusters.~\cite{Han} In contrast, Nayak {\em et al.} predicted the ground state magnetic coupling to be ferromagnetic for (MnO)$_x$ ($x$=2--6 and 9).~\cite{Jena1,Jena2} We argue that a restricted potential energy search is responsible for such wrong prediction.~\cite{Jena1,Jena2}  {We find that the dimensionality of the stoichiometric clusters in their respective ground state can be explained in terms of total hybridization. For all the clusters, the calculated $\mathcal{H}$ is higher for the ground state, which also explains the observed 2D$\rightarrow$3D structural transition with increasing MnO-units (Table~\ref{tab1}). This also explains the  predicted magnetic structure for the ground state -- AFM is favorable over the corresponding FM solution.~\cite{MnO} It is important to note that irrespective of the cluster size, the Mn$\rightarrow$O charge transfer remain constant ($\sim$1.3$e$). }

\subsection{Mn dimer: {Effect of electron correlation}}

\begin{figure}[t]
\includegraphics[scale=0.45]{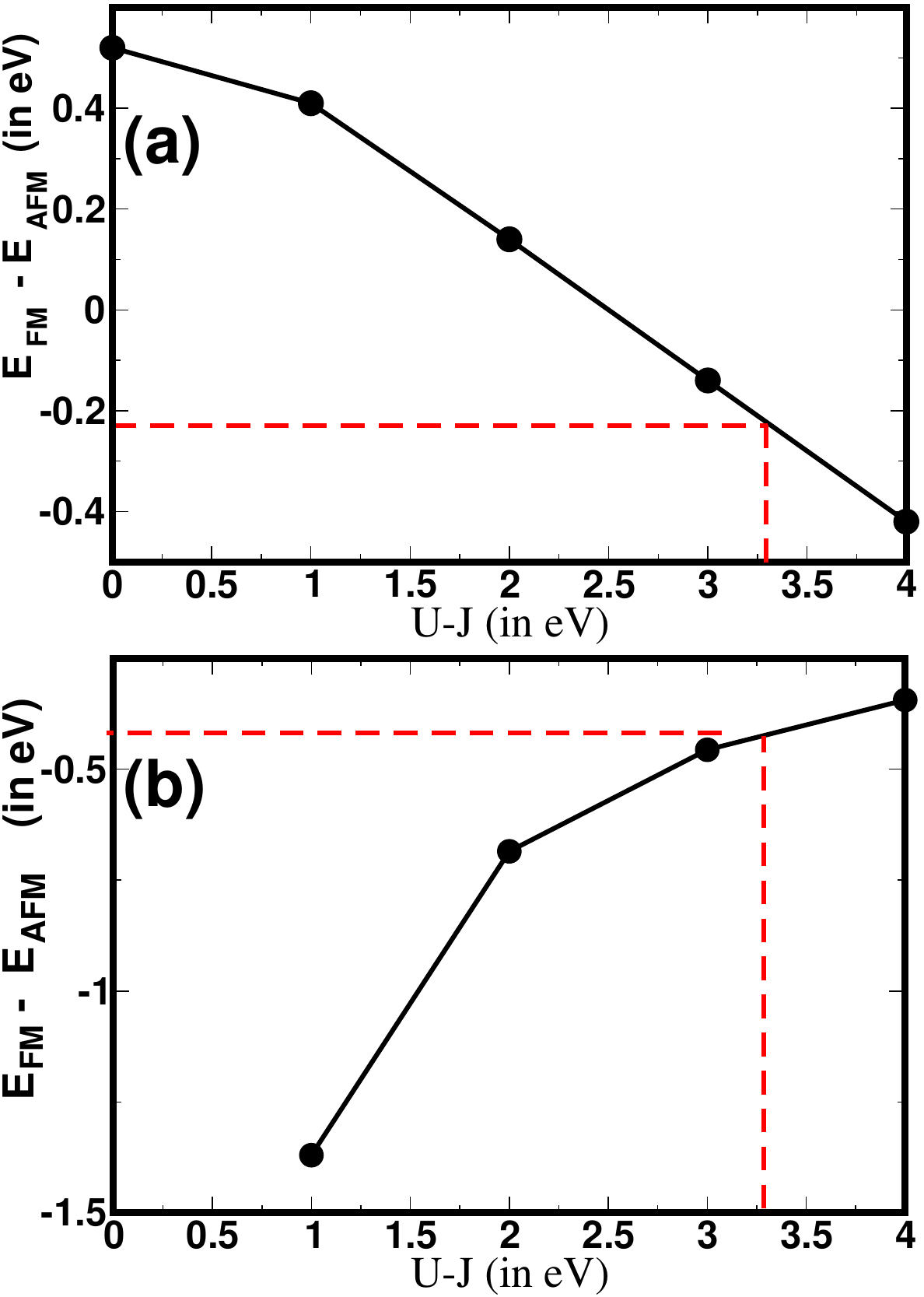}
\caption{(Color online) Effect of electron correlation in Mn-dimer. (a) Calculated energy difference between the FM and AFM states within the DFT$+U$ approach. We observe FM to AFM crossover due to the increase in on-site Coulomb interaction. Within this approach, $U\sim$ 3.25 eV reproduces the energy difference calculated using the PBE0 hybrid functional. (b) { Calculated energy difference between the FM and AFM states within the model Hamiltonian approach using Eqn.~\ref{jij}, where the hopping parameters are calculated using the maximally localized Wannier function.} The dotted line indicates the energy difference between the FM and AFM solutions for $U\sim$ 3.25 eV.}
\label{MnO-U}
\end{figure}

Before we discuss the off-stoichiometric MnO clusters, here we discuss the debated Mn-dimer. The magnetic coupling for Mn$_2$ is predicted to be ferromagnetic within DFT calculations using conventional exchange-correlation functional, and the closest AFM state is found to be 0.52 eV higher in energy.~\cite{Kabir2006} However, this result within conventional DFT is in contrast with the predictions from resonance Raman spectroscopy,~\cite{Moskovits} and electron spin resonance measurements,~\cite{Baumann} which argue the Mn-Mn coupling to be antiferromagnetic. {In this regard, we revisit the Mn-dimer to include the strong electron correlation, and perform DFT calculations including the on-site Coulomb interaction (DFT+$U$). Indeed, we find that electron correlation affects the Mn-Mn coupling, which becomes antiferromagnetic for $U\geqslant$ 2.5 eV [Fig.~\ref{MnO-U}(a)].} We reconfirm this observation using hybrid exchange-correlation functional PBE0, where we find an AFM ground state, which is 0.23 eV lower in energy compared to the corresponding FM solution. {Comparing the DFT+$U$ and PBE0 calculations, we find that $U\sim$ 3.25 eV reproduces the same stability for the AFM ground state over the FM solution [Fig.~\ref{MnO-U}(a)].} The Mn-Mn bond length for the dimer within conventional PBE is much smaller (2.58 \AA) than the same calculated using hybrid PBE0 functional (3.11 \AA).  { This is in agreement with the experimental predictions (3.13--3.4 \AA).~\cite{Moskovits, Baumann} Although, the local density approximation based calculation predicted AFM ground state,~\cite{Mejia-Lopez} the calculated bond length was found to be much smaller (2.89 \AA) than the experimentally predicted range.~\cite{Moskovits, Baumann} Moreover, in this calculation, the magnetic ground state becomes ferromagnetic in the experimental bond length regime ($\geqslant$3.06 \AA),~\cite{Mejia-Lopez} which contradicts the experimental results. In contrast, the present PBE0 calculation always predicts an AFM solution in the experimental bond length regime. Thus, Mn-dimer can not be described within the conventional DFT, and one needs to incorporate strong electron correlation to correctly reproduce the experimental results.~\cite{Moskovits, Baumann}} Now it would be interesting to see if the same is true for other small Mn-clusters. Interestingly, we find that for Mn$_x$ ($x\leqslant$ 5) clusters, both conventional PBE, and PBE0 functionals predict ferromagnetic ground state. Thus, the electron correlation may not be crucial for these clusters. 

Next, we analyse the Mn-Mn direct exchange interaction within model Hamiltonian approach, where we neglect the 4$s$ electrons. In the absence of electron hopping, the ground state will have five singly occupied $d$-orbitals.~\cite{Koch} This approach is valid for any system, where the interacting Mn atoms have five singly occupied $d$-orbitals, and we will show later that this is indeed the case for stoichiometric (MnO)$_x$ clusters. We consider a tight binding Hamiltonian $\mathbf{H}$ in the limit where the on-site Hubbard correlation $U$ is much larger compared to the hopping integral, ($U\gg t$), and thus, a perturbative treatment of the $\mathbf{H}$ is possible. The Hamiltonian $\mathbf{H}$ can be written as,
\begin{equation}
\mathbf{H}=\mathbf{H}_U+\lambda\mathbf{H}_t,
\end{equation}
where $\lambda$ is a continuous real parameter. This parameter is introduced to keep track of the number of times the perturbation enters. At the end of the calculation we may set $\lambda \rightarrow 1$ to get back the full-strength case.~\cite{Sakurai} Here the $\mathbf{H}_U$ is given by,~\cite{Pavarini}
\begin{eqnarray}
\mathbf{H}_U &=& U\sum_{i=1,2}\sum_{m=1,..5}n_{i,m,\sigma}n_{i,m,-\sigma}\nonumber \\ 
&+& \frac{1}{2}\sum_{i,\sigma,\sigma'}\sum_{m\neq m'}(U-2J_{H}-J_{H}\delta_{\sigma,\sigma'})n_{i,m,\sigma}n_{i,m',\sigma'} \nonumber \\
&+& \sum_{i=1,2}\sum_{m=1,...5}E_{m}^{i},
\label{H1}
\end{eqnarray}
and the hopping Hamiltonian $\mathbf{H}_t$, which acts as the perturbation in this treatment is, 
\begin{eqnarray}
\mathbf{H}_t=\sum_{m,m'=1,5}t^{1,2}_{m,m'}c^{+}_{1,m}c_{2,m'}+t^{2,1}_{m,m'}c^{+}_{2,m}c_{1,m'},
\label{H0}
\end{eqnarray}
where 1 and 2 are the indices for two Mn sites. The indices $m$ and $m'$ run over the five Mn-$d$ orbitals. $c^{\dagger}_{im\sigma}c_{im\sigma}$=$n_{im\sigma}$, and  $\{  c^{\dagger}_{im\sigma}, c_{im\sigma}\} $ are the electron creation and annihilation operators for the orbital $m$ with spin $ \sigma $ for the $i$-th site. $J_{H}$ is the Hund's coupling constant, and $E_{m}^{i}$ is the local on-site energy at the $i$-th site for the orbital $m$. The hopping integral $t^{i,j}_{m,m'}$ represents the hopping between $m$-th orbital of site $i$ and $m'$-th orbital of site $j$, { which are calculated using the maximally localized Wannier function.} 

The Heisenberg exchange interaction is bilinear in spin, and for Mn$_2$ dimer, 
\begin{equation}
\pazocal{H}=J\mathbf{S_{1}}\cdot\mathbf{S_{2}},
\end{equation}
where $J$ is the exchange coupling, and $\mathbf{S_{i}}$ is the localized spin at the $i$-th site. It can be shown from Eqn.~\ref{H1} and Eqn.~\ref{H0} that the direct exchange interaction is, 
 \begin{eqnarray}
J=\frac{E_{\rm FM}-E_{\rm AFM}}{2S^2} \propto \frac{2\sum_{m,m'=1}^{5}|t_{m,m'}^{1,2}|^2}{U+4J_{H}} 
\label{jij}
\end{eqnarray}

Thus, the direct exchange mechanism within this model makes Mn-Mn coupling antiferromagnetic, for all values of $U$ studied here [Fig\ref{MnO-U} (b)]. {This picture of direct exchange is valid for two interacting Mn atoms with localized 3$d$ electrons. We calculate the energy difference between the AFM ground state and the excited FM solution with varied on-site Coulomb $U$, which decreases with increasing $U$. For $U\sim$ 3.25 eV, the interpolated energy difference is $-$0.43 eV, which is slightly larger than the one calculated within the PBE0 hybrid functional. Thus, all these calculations confirm the necessity of on-site Coulomb interaction to describe Mn-dimer that is consistent with experimental predictions,~\cite{Baumann,Moskovits}  which is not possible within the conventional DFT calculations.}
   
\begin{figure*}[t]
\includegraphics[scale=0.5]{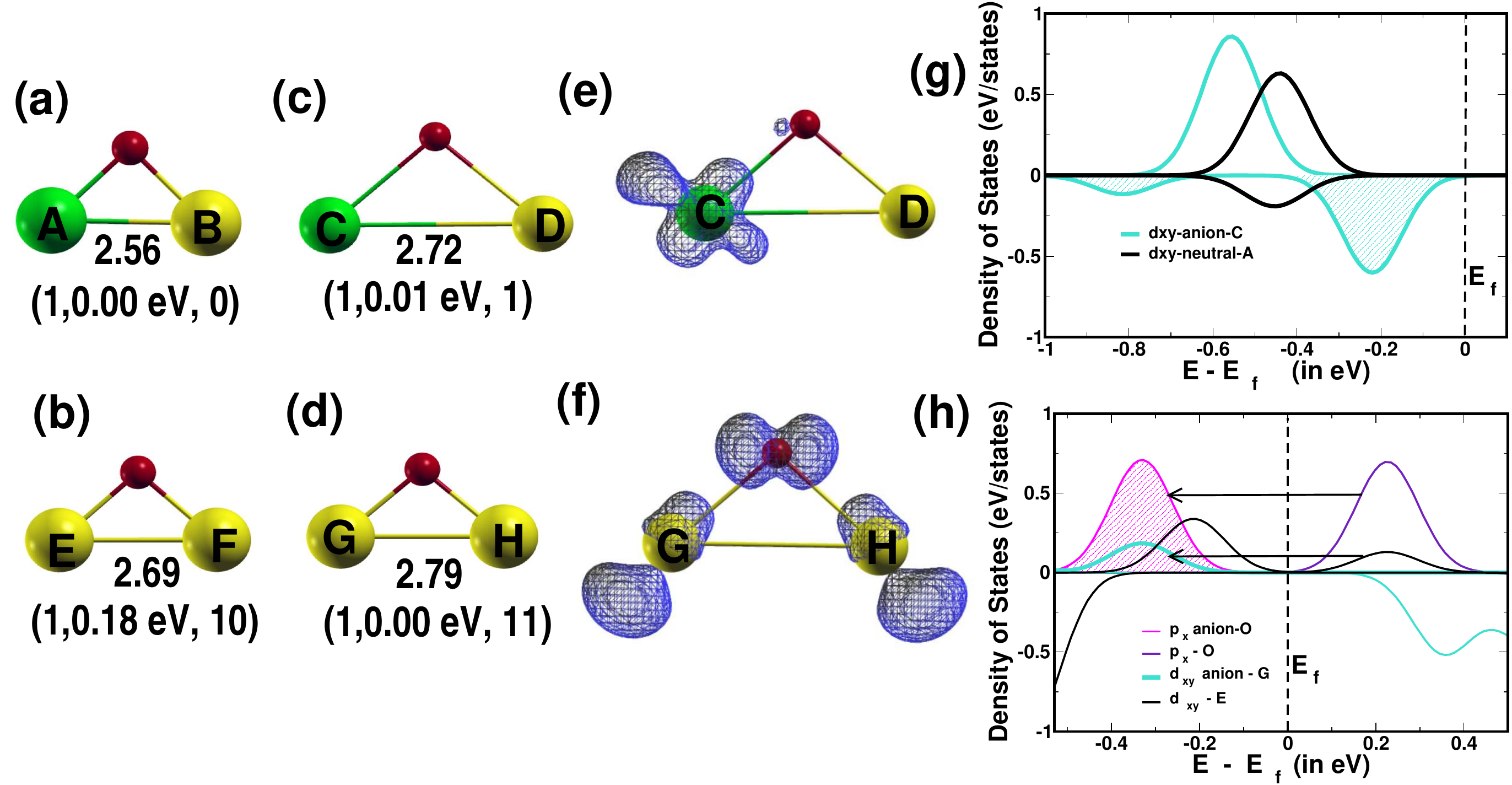}
\caption{\label{composite}(Color online) Neutral Mn$_2$O [(a) and (b)] and anionic Mn$_2$O$^-$ [(c) and (d)] clusters that are calculated using the PBE exchange-correlation functional. While the extra electron is localized on a particular Mn for the AFM case [(e)], the extra electron is distributed over the cluster [(f)] for the FM solution, which increases the Mn-O hybridization. (g) Calculated atom projected density of states for the AFM solution in neutral and anionic cases, which show the localization of the added electron for the anion. (h) Atom projected density of states show an increase in Mn-O hybridization for the anionic FM solution compared to the neutral Mn$_2$O. }
\end{figure*}    

\subsection{Off-stoichiometric MnO clusters}

We start our discussion with neutral Mn$_2$O and anionic Mn$_2$O$^-$ clusters {to study if the overall charged state of the cluster alters the Mn-Mn magnetic coupling.} Although the magnetic ground state of Mn$_2$ crucially depends on the choice of exchange-correlation functional, the neutral Mn$_2$O is found to be antiferromagnetic within both PBE, and hybrid PBE0 functional. However, the energy difference between the AFM ground state and the excited FM state is found to be very different, 0.18 and 0.64 eV for PBE [Fig.~\ref{composite} (a) and (b)] and PBE0 functionals, respectively. Thus, hybrid functional stabilizes the AFM structure more than for the PBE case. In contrast, the addition of one extra electron to neutral Mn$_2$O completely changes the magnetic ground state -- the Mn-Mn coupling becomes ferromagnetic for Mn$_2$O$^{-}$ anion, which is in agreement with the previous theoretical and experimental observations.~\cite{Jena4} However, the energy difference is very small 0.01 eV between the AFM and FM solutions [Fig.~\ref{composite} (c) and (d)], which is calculated to be 0.19 eV within the hybrid PBE0 functional. 

In addition to the Mn-Mn direct exchange, the presence of oxygen in the MnO clusters makes the magnetic coupling more complex due to possible superexchange interaction mediated via oxygen. For neutral AFM coupled Mn$_2$O cluster,  the up (down) spin channel is completely filled for Mn1 (Mn2), while the other spin channel is completely empty. Thus, the virtual hopping of the electrons from one Mn to the other Mn is possible only for the AFM structure, while such virtual hopping is restricted for the FM solution. Thus, the AFM structure becomes the ground state for the neutral cluster. However, the situation for Mn$_2$O$^-$ is very different, and the calculated charge density for the added electron is shown in Fig.~\ref{composite} (e) and (f) for the AFM and FM configurations, respectively. It is evident that for the AFM case, the extra electron is localized on a particular Mn atom [Fig.~\ref{composite} (e)], and populates the down channel  [Fig.~\ref{composite} (g)], and thus the available channels for virtual spin hopping decreases. In contrast, for the FM solution, the extra electron participates in bonding and increases the $p_{\rm O}-d_{\rm Mn}$ hybridization [Fig.~\ref{composite} (f) and (h)]. {Thus, the FM configuration becomes the ground state for Mn$_2$O$^-$, which is otherwise AFM for its neutral counterpart.}

Next, we turn our attention to Mn$_x$O$_y$ clusters ($x$=3--4 and $y \leqslant x$) to investigate the evolution of magnetic structure due to monotonic increase in the number of oxygen. It has been already predicted that the pure Mn$_3$ and Mn$_4$ are ferromagnetic,~\cite{Kabir2006} while both the stoichiometric (MnO)$_3$ and (MnO)$_4$ clusters are found to have AFM ground state.~\cite{MnO}  Thus, it would be interesting to study the evolution of magnetism due to chemical doping, { and indeed we find that both cluster geometry and the magnetic coupling are strongly influenced by oxygen.} The calculated binding energy, energy difference between the most stable AFM and FM solutions, hybridization, and the corresponding Mn$\rightarrow$O charge transfer are tabulated in Table \ref{tab2}.

\begin{figure}[t]
\includegraphics[scale=0.5]{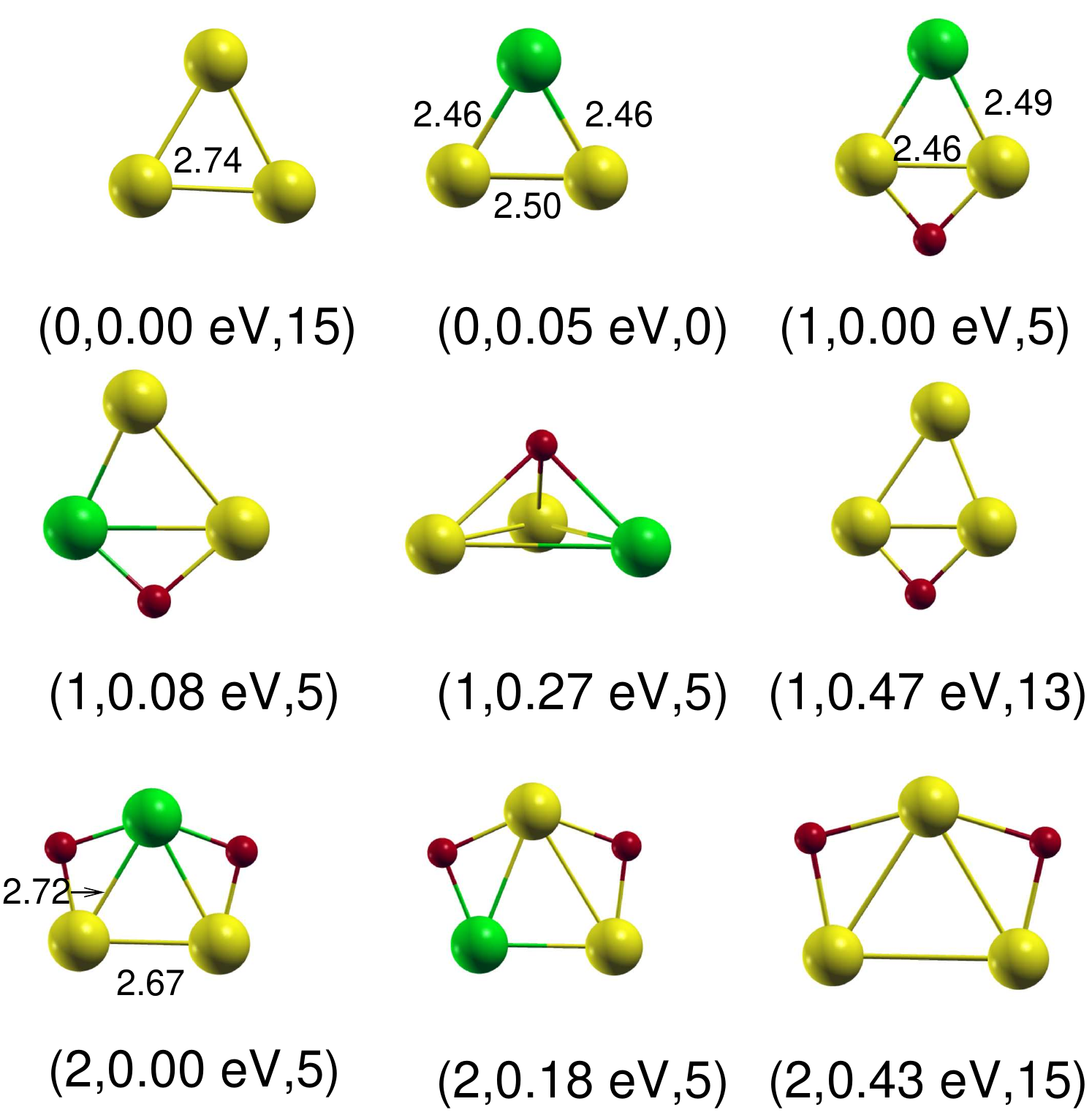}
\caption{\label{Mn3Oy}(Color online) Ground state and the significant isomers for off-stoichiometric Mn$_3$O$_y$ ($y$=0--2) clusters. The numbers in the parenthesis indicate the number of oxygen atoms $y$ in the cluster, the energy relative to the ground state, and the total magnetic moment of the cluster. The Mn-Mn distances for the ground state are shown. Unlike the pure Mn$_3$ and stoichiometric (MnO)$_3$ clusters, for these clusters the FM coupled Mn atoms are closer in space than the AFM coupled ones.}
\end{figure}

\begin{table}[b]
\caption{\label{tab2}Binding energy per atom, energy difference between the most stable AFM and FM solutions ($\Delta E = E_{\rm AFM} - E_{\rm FM}$) for the off-stoichiometric MnO clusters. The total hybridization index $\mathcal{H}_{\rm tot}$ ($p_{\rm O}-d_{\rm Mn}$ hybridization index  $\mathcal{H}_{pd}$), and the total Mn$\rightarrow$O charge transfer are also shown.}
\begin{tabular}{lcrccc}
\hline
\hline
Cluster & $E_B$ &  $\Delta E$ &  \ \ \ &  $\mathcal{H}_{\rm tot}$ ($\mathcal{H}_{pd}$)    & Mn$\rightarrow$O \\
        &  (eV/atom) & (eV) & &  & CT ($e$)\\
\hline
\hline
Mn$_3$ & 0.82 & 0.05 && $-$ & $-$\\
Mn$_3$O & 2.47 & -0.47& & 1.37 (0.99) & 1.0\\
Mn$_3$O$_2$ & 3.50 & -0.43 & & 3.91 (2.98) & 1.2\\
Mn$_3$O$_3$ & 4.13 & -0.42 & & 4.86 (3.84) &1.2\\
Mn$_4$ & 1.18 & 0.08 & & $-$ & $-$ \\
Mn$_4$O & 2.36 & -0.59 & & 2.47 (1.45)& 1.0\\
Mn$_4$O$_2$ & 3.28 & -0.95 &  & 4.19 (3.05)& 1.2\\
Mn$_4$O$_3$ & 3.84 & -0.74 & & 4.67 (4.09) &1.2\\
Mn$_4$O$_4$ & 4.29 & -0.73 & & 5.68 (4.30)&1.2\\
\hline
\end{tabular}
\end{table}

As we have mentioned earlier that the ground state for the pure Mn$_3$ is ferromagnetic, and the most stable AFM solution with 5$\mu_B$ moment lies only 50 meV higher in energy (Fig.~\ref{Mn3Oy}).  In contrast, addition of a single oxygen to the pure Mn$_3$ trimer makes the neutral Mn$_3$O cluster antiferromagnetic, and the corresponding geometry remains two-dimensional (Fig.~\ref{Mn3Oy}).  The most stable FM isomer is found to be quite high in energy (0.47 eV). {In contrast, a recent calculation predicted a three-dimensional geometry for the neutral Mn$_3$O cluster, which was motivated by their three dimensional Mn$_3$O$^-$ anion.~\cite{MnO-recent}  However, we find such structure to be 0.27 eV higher in energy (Fig.~\ref{Mn3Oy}).} The magnetic ground state remains AFM on further addition of O as Mn$_3$O$_2$ and (MnO)$_3$ clusters are found to be antiferromagnetic. The corresponding energy difference $\Delta E$ remains $\sim$ $-$0.43 eV for these clusters (Table \ref{tab2}).  

\begin{figure}[t]
\includegraphics[scale=0.5]{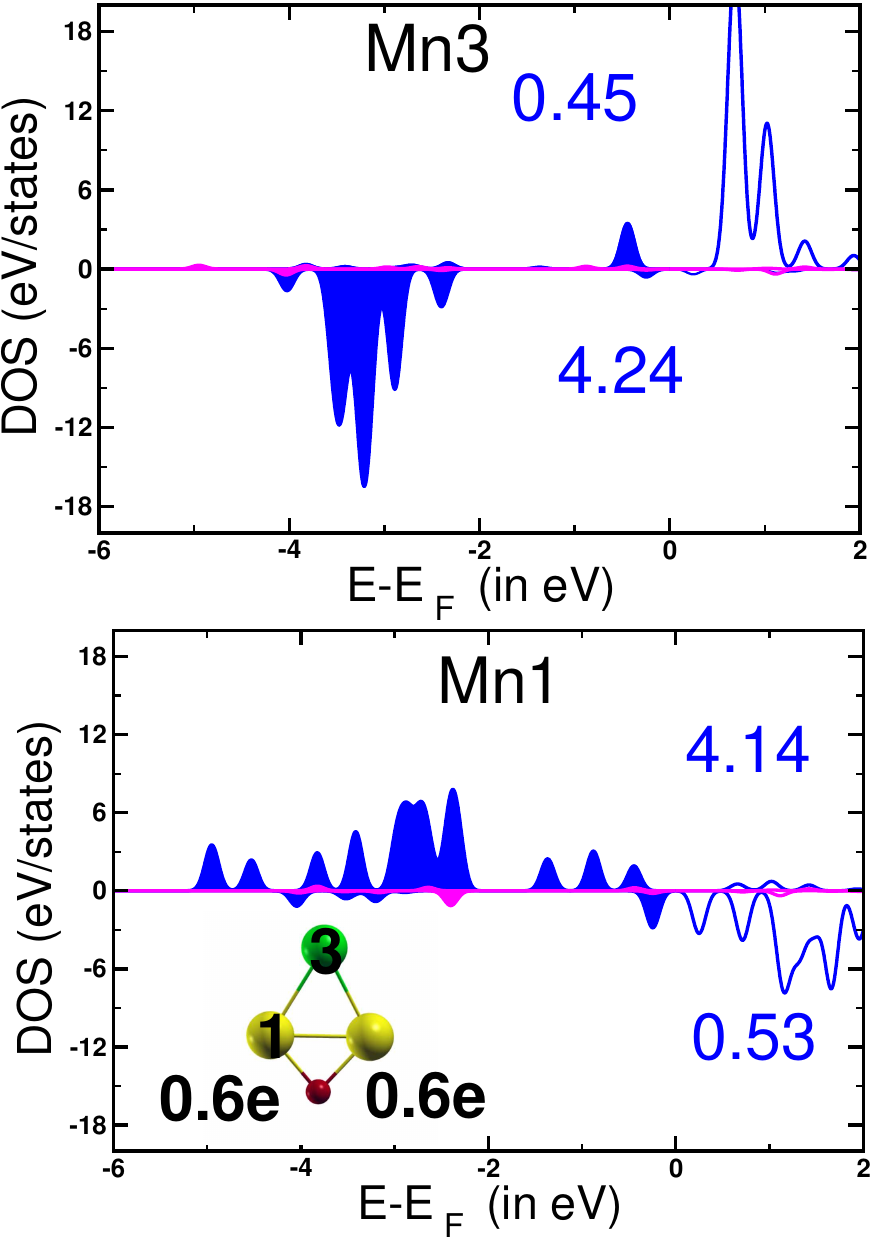}
\caption{\label{chgtransf}(Color online) The total amount of charge that is transferred to an oxygen atom remains constant irrespective of the stoichiometry of the cluster. Thus, for an off-stoichiometric cluster (shown for Mn$_3$O) all the Mn atoms are not in the same charge state. The density of states of the $d$-orbitals (blue), corresponding to the structure shown in the inset, show that the polarization of the $d$-orbital (indicated by the number in blue) reduces in oxygen environment.}
\end{figure}

We calculate the Mn$\rightarrow$O charge transfer in these clusters using Bader analysis.~\cite{bader} We see that for both stoichiometric and off-stoichiometric clusters, and irrespective of the size and composition, the oxygen atoms receive a total of $\sim$ 1.2$e$ charge from the neighbouring Mn atoms (Table \ref{tab1} and \ref{tab2}). Thus, while for the stoichiometric clusters all the Mn atoms are in the same charged state, this is not the case for off-stoichiometric clusters due to unequal number of Mn and O atoms (Fig.~\ref{chgtransf}). We find that the density of states for the $d$-electrons are relatively more delocalized, and less polarized for the Mn atom from which charge is transferred. However, the net occupancy of the Mn-$d$ level remains nearly half-filled irrespective of coordination. Analysis of superexchange interaction between two cations, with half-filled $d$ shell, and $\sim$ 90$^{\circ}$ cation-O-cation bond angle, within the existing theories is difficult.~\cite{Kanamori} Thus, we devise an alternate route to gain a quantitative insight into the nature of the magnetic coupling in these cluster.

\begin{figure*}[t]
\includegraphics[scale=0.6]{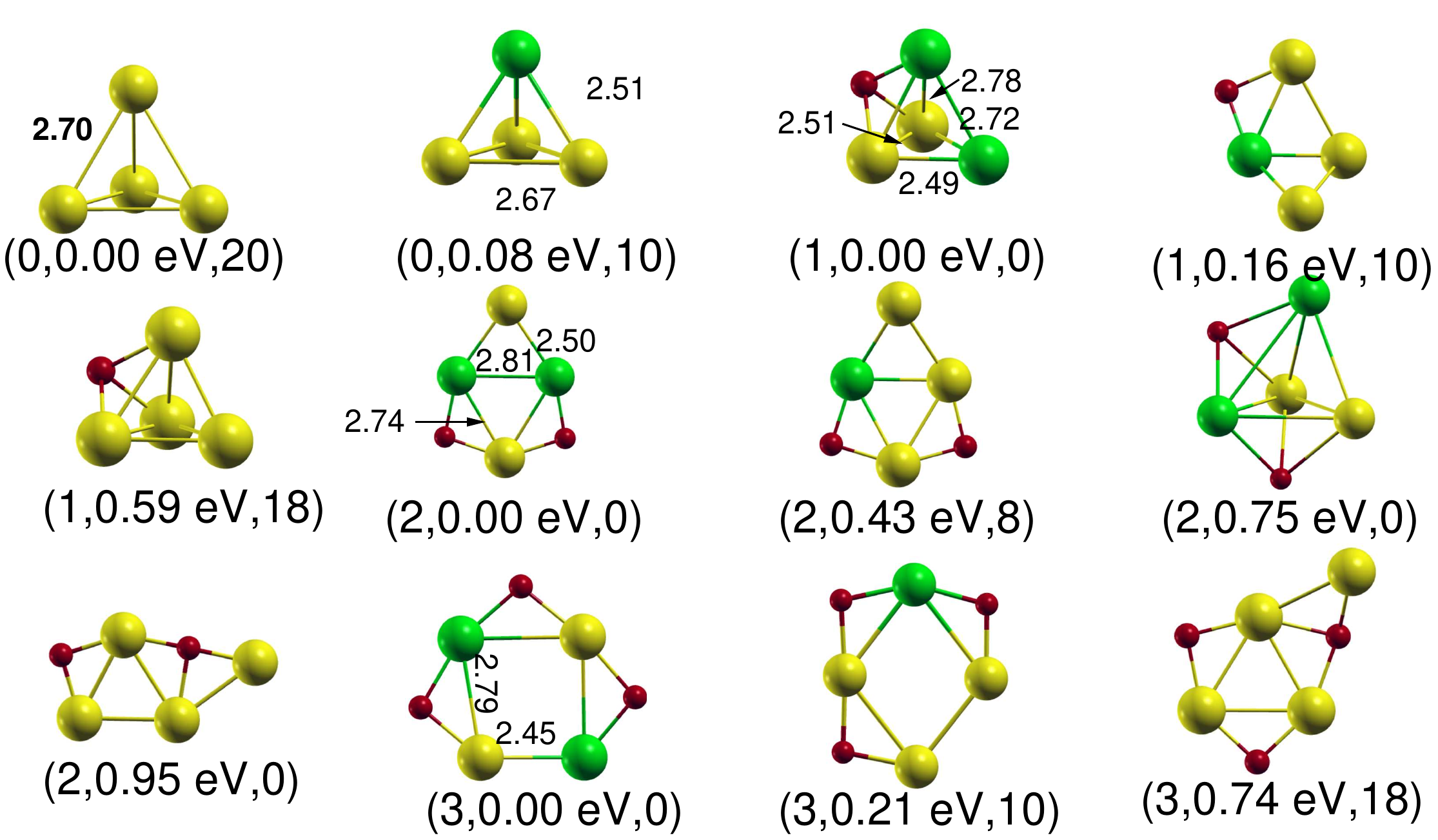}
\caption{\label{Mn4Oy}(Color online) Ground state and the significant isomers for the off-stoichiometric Mn$_4$O$_y$ ($y$=0--3) clusters. The numbers in the parenthesis indicate the number of oxygen atoms $y$, the energy difference with the corresponding ground state, and the total magnetic moment of the cluster.}
\end{figure*}

As we have discussed earlier that the magnetic structure in these MnO clusters is rather complex, where the Mn-Mn direct exchange compete with the superexchange mediated via oxygen. For example, here we calculate the hopping between the Mn atoms in (MnO)$_2$ using the maximally localized Wannier functions. In this regard, we calculate both direct hopping between the two Mn atoms, and the effective hopping including the one mediated by the oxygen. Specifically, we calculate $\sum_{m,m'=1}^{5}|t^{1,2}_{m,m'}|^2$, where $m$ and $m'$ indices represent five $d$-orbitals, which roughly depicts the probability of an electron to jump from one site to the other.  Considering only direct hopping, we calculate this quantity to be 1.4 eV$^2$, which becomes 2.5~eV$^2$ once we include the hopping mediated via oxygen. This reflects that although the direct hopping is predominant, the hopping mediated via oxygen is also substantial, which leads to superexchange interaction. {This demonstrates that both direct exchange and superexchange play important role in determining the magnetic structure in these clusters.}

For pure Mn-clusters, average bond length between the AFM coupled atoms is 3-8\% shorter than that of the FM coupled atoms, which was earlier predicted by one us, and was explained in terms of Pauli exclusion principle.~\cite{Kabir2006} Note that in pure clusters the magnetic structure is dictated only by the direct exchange interaction.  In contrast to the pure Mn-clusters, for Mn$_3$O and Mn$_3$O$_2$, the FM coupled Mn atoms are closer in space than those coupled antiferomagnetically. This further substantiates the claim that direct exchange is not the primary mechanism of magnetism in MnO clusters. Due to unequal number of Mn and O atoms for off-stoichiometric clusters, the charge state of all the Mn atoms are not the same. Moreover, we find that the Mn atoms which have same charged state are coupled ferromagnetically. However, the situation is quite different for stoichiometric clusters, where all the Mn atoms are in the same charge state, as each Mn atoms looses same amount of charge to the neighbouring O atoms. Thus, for the stoichiometric (MnO)$_3$ cluster, we get a similar picture as in pure Mn$_3$: FM coupled Mn atoms are far apart (2.82 \AA) compared to the AFM coupled ones (2.62 \AA).

Next, we study the magnetic evolution in Mn$_4$O$_y$ clusters with increasing oxygen concentration (Fig.~\ref{Mn4Oy}). While the pure Mn$_4$ cluster is ferromagnetic,~\cite{Kabir2006} addition of single oxygen makes Mn$_4$O to be antiferromagnetic. A two-dimensional ferrimagnetic structure (10$\mu_B$) is found to be 0.16 eV higher in energy. { In contrast with the present results, a three dimensional ferrimagnetic structure with 10$\mu_B$ moment was reported to be the ground state in a recent study.~\cite{MnO-recent} We find this structure to be 0.19 eV higher in energy than the AFM ground state, and this structure also has non-degenerate isomag, which is 0.22 eV higher in energy. Although these results on three dimensional structure with 10$\mu_B$ moment agree with the earlier predictions,~\cite{MnO-recent} we predict AFM ground state. We argue that this discrepancy is due to the fact that their cluster geometry/magnetism was motivated by the results on the corresponding anionic clusters. Thus, unlike the present calculations, the potential energy search for neutral clusters was biased, which assume the overall charged state does not influence the geometric and/or magnetic structure. This assumption might not be true, and indeed we find that the charged state strongly influence both geometric and magnetic structures, which we will discuss later.}

The AFM structure remains the ground state for further increase in oxygen concentration in the cluster (Table~\ref{tab2}). For Mn$_4$O$y$ clusters, in addition to FM $\rightarrow$ AFM transition, we also observe 3D$\rightarrow$2D structural transition (Fig.~\ref{Mn4Oy}) at $y$=2, and the ground state remains 2D for further increase in oxygen concentration. These can be explained in terms of Mn-O hybridization.~\cite{MnO}  For these clusters, the closely lying isomers are all found to couple either antiferromagnetically or ferrimagnetically. The most stable FM structures are found to be very high in energy, 0.59, 0.95 and 0.74 eV for  Mn$_4$O, Mn$_4$O$_2$ and Mn$_4$O$_3$, respectively. {It is interesting to note that the binding energy increases monotonically with increasing oxygen concentration for both Mn$_3$O$_y$ and Mn$_4$O$_y$ clusters (Table~\ref{tab2}). This is due to the monotonic increase in $p_{\rm O}-d_{\rm Mn}$ hybridization with oxygen concentration.}

{It would be interesting to study the effect of adding an extra electron on the geometric and magnetic structure for these clusters.} As we have mentioned earlier that the neutral Mn$_3$O is two-dimensional. In contrast, the Mn$_3$O$^-$ anion is found to be three-dimensional, where the oxygen sits on top of a planer Mn$_3$, and bonded with all the Mn-atoms. This is in agreement with the experimental prediction.~\cite{MnO-recent}  Although, both the 3D and 2D structures have ferrimagnetic Mn coupling with 6$\mu_B$ moment, the 2D anionic isomer lies 0.22 eV higher in energy.  Bader analysis indicates that the extra electron is localized on a particular Mn atom for the 2D isomer, which spreads over the cluster in the 3D structure, which increases the $p_{\rm O}-d_{\rm Mn}$ hybridization and consequently becomes the ground state.  The magnetism in Mn$_4$O$^-$ anion is very interesting compared to the neutral counterpart. The corresponding ground state is found to be ferrimagnetic with large moment (11$\mu_B$) compared to the completely compensated antiferromagnetic structure for the neutral case (Fig.~\ref{Mn4Oy}).  The AFM state is found to be slightly higher (0.1 eV) in energy. { These results on anionic clusters are in agreement with the previous calculations.~\cite{MnO-recent} However, as we have discussed earlier, the previous results on the neutral counterpart do not agree with the present calculations. This is due to the fact that the potential energy surface scan for neutral clusters was largely biased by the results for the anionic counterparts, and thus the previous calculations failed to predict the correct ground state for neutral Mn$_3$O and Mn$_4$O clusters.~\cite{MnO-recent} On the other hand, the present ground state search is unbiased, and indeed we observe that the overall charged state of the cluster strongly influence the ground state geometry and/or the magnetic structure of the cluster. These results open up a possible way to manipulate the intrinsic magnetism in small MnO clusters on a surface.}

\subsection{\label{sec:n13} Vertical Displacement Energy}

We calculate the vertical displacement energy (VDE) as the energy difference between the anion and neutral cluster;  ${\rm VDE}=E^{n+1}-E^{n}$, 
where $n$ is the number of electrons in the cluster, and $E^{n+1}$ and $E^{n}$ are the binding energies of the anionic, and corresponding neutral clusters. In these calculations, the geometry of the neutral cluster is fixed at the corresponding anionic structure. If $S$ is the ground state moment of the anion, the corresponding neutral cluster has either $S+1$ or $S-1$ magnetic moment. The calculated VDEs using both PBE and PBE0 hybrid functionals are tabulated in Table~\ref{table3}, which are in good agreement with the photoelectron spectroscopy measurements.~\cite{MnO-recent} Good agreement between the calculated PBE and experimental results indicate that the PBE exchange-correlation functional may be enough to describe the MnO clusters. 

\begin{table}[t]
\caption{\label{table3}Calculated VDE for Mn$_x$O clusters ($x$=2, 3 and 4) for $S \rightarrow S\pm 1$ transition using PAW-PBE and PBE0 functionals.  Calculated VDEs are compared with previous theoretical and experimental studies.~\cite{MnO-recent, Jena4} }
\begin{tabular}{llcccc}
\hline
\hline
Cluster & VDE  & PAW  & Gaussian& PAW & Exp.\\
        &   $S \rightarrow S \pm 1$                       &   (PBE)       &   (PBE) & (PBE0) & \\
        &  ($\mu_{B}$)  & (eV)  & (eV) & (eV) & (eV) \\
\hline 
\hline 
Mn$_2$O & 11 $\rightarrow$ 12 & 2.06 & 2.00 & 1.59 & 1.56, 1.75,\\
        & 11$\rightarrow$ 10 & 1.38 & 1.35 & 1.71 &  2.04 [Ref.~\onlinecite{footnote}] \\
Mn$_3$O & 6 $\rightarrow$ 7 & 1.86 & 1.93 & 1.92   & 2.09 \\
        & 6 $\rightarrow$ 5 & 1.30 & 1.62 & 1.45   & 1.68 \\
Mn$_4$O & 11 $\rightarrow$ 12 & 1.99 & 2.48 & 1.85 &  2.53  \\
        & 11 $\rightarrow$ 10 & 1.58 & 1.92 & 1.64 &  2.05 \\
\hline
\end{tabular}
\end{table}

\section{Summary and conclusions}

{
Using density functional theory, we study the evolution of geometric and magnetic structure in MnO-clusters depending on the oxygen concentration, and the charged state. In the present study, we have also considered the enhanced electron correlation, which we find to be critical in predicting the correct experimental ground state for the Mn-dimer. In this regard, we find that the conventional DFT calculations fail to reproduce the experimental results. In contrast, the present calculations using the strong electron correlation within DFT+$U$, hybrid exchange-correlation functional, and model Hamiltonian based calculations predict the Mn-Mn direct exchange to be antiferromagnetic, which is in agreement with the experimental observations.~\cite{Moskovits, Baumann}  However, the situation for small Mn-clusters is very different, where we find that both conventional PBE and hybrid PBE0 exchange-correlation functionals predict the same magnetic ground states. This may indicate that the inclusion of strong correlation may not be necessary for these clusters. 

In comparison to the pure Mn-clusters, the magnetic structure in MnO-clusters is complex due to competing direct exchange and superexchange interactions. In these MnO-clusters, we find the oxygen mediated electron hopping to be substantial, and that plays a crucial role in determining the magnetic structure. Thus, in general, the Mn-Mn coupling is antiferromagnetic due to the presence of oxygen, which is otherwise ferromagnetic for small Mn-clusters.  We also observe a 3D$\rightarrow$2D structural transition due to oxygen doping, which can be explained in terms of $p_{\rm O}-d_{\rm Mn}$ hybridization.  Interestingly, the charged state of the cluster strongly influences the geometric and/or magnetic structure of the MnO cluster, which is explained with the help of $p_{\rm O}-d_{\rm Mn}$ hybridization, and the distribution of the added electron in anionic MnO-clusters. Although, the results on the anionic MnO-clusters are in agreement with the previous theoretical study,~\cite{MnO-recent} some of the results on neutral MnO clusters are in contrast. This discrepancy is due to the biased ground state search for the neutral clusters, which are motivated by the corresponding anionic clusters.~\cite{MnO-recent} Calculated vertical displacement energies using both conventional PBE and PBE0 hybrid functionals are in good agreement with the available experimental results,~\cite{MnO-recent} which indicate that the conventional exchange-correlation functional may be enough to describe these small MnO-clusters. We hope that the present electronic structure analysis provides a microscopic understanding to the complex magnetic structure in MnO-clusters. Further, the present calculation on the anionic clusters indicate a possible way to manipulate the intrinsic magnetism via carrier doping, and will motivate investigations of such clusters on substrate.  
}

\acknowledgements
MK acknowledges grant from the Department of Science and Technology, India under Ramanujan Fellowship. CA and BS acknowledge financial support from Carl Tryggers Stiftelse (grant no. CTS 12:419). BS acknowledges VR/SIDA for financial support. Supercomputing facilities allocated by Swedish National Infrastructure for Computing (SNIC) is gratefully acknowledged. Some of the calculations were done using the supercomputing facility at the Inter University Accelerator Centre, Delhi.


\begin{thebibliography}{99}
\bibitem{footnote}Significant peaks in the experimental photoelectron spectra are found to be at 1.56, 1.75, and 2.04 eV, and proposed to be a result of transition from two nearly degenerate anionic isomers (ferromagnetic and antiferromagnetic).
\end{thebibliography}
\end{document}